\documentclass{elsart}

\usepackage{epsf}

\begin{document}
\begin{frontmatter}
\title{Faster Background Determination\\
    {\large \-- a method for gaining time coverage and\\
    flux measurement accuracy with Cherenkov telescopes}}

\author{Dirk Petry}
\address{Joint Center for Astrophysics, University of
Maryland Baltimore County,\\ 1000 Hilltop Circle, Baltimore,
MD 21250,\\ and NASA/GSFC, Code 661, Greenbelt, MD 20771
}

\maketitle

\begin{abstract}
An improved way of taking off-source data for background determination
in Cherenkov telescope observations is proposed. Generalizing
the traditional concept of taking on-source/off-source observations
of equal duration (e.g. 30 minutes ON followed by 30 minutes OFF),
{\it Faster Background Determination} (FBD) permits an
off-source observation with the same zenith angle distribution
as the on-source observation to be obtained within less time. The method
permits the on-source observation time to be maximized 
without compromising the quality of the background determination.
It also increases the signal significance for strong sources.
The only modification necessary in the data acquisition is
a small change to the tracking algorithm.
The only modification necessary in the data analysis is to introduce
a time normalization which does not increase the systematic errors.
The method could become the normal observing mode for
Cherenkov telescopes when observing strong sources.
\end{abstract}

\end{frontmatter}

\section{Introduction}

Since their first successful application in the late 1980s
\cite{weekes89}, Imaging Atmospheric Cherenkov Telescopes (CTs) 
have developed rapidly from pioneer instruments to 
precision observatories for high-energy gamma radiation
with a large user community.
With four major new observatories under construction
(CANGAROO III \cite{mori}, HESS \cite{hofmann}, MAGIC
\cite{lorenz}, VERITAS \cite{weekes2}), it is justified to revisit
and optimise the standard CT data taking methods in terms of achieving the
best possible scientific output given the limited observation time.  

CTs can only observe at night (ideally moonless) and during good weather conditions.
These constraints typically lead to a total yearly observation
time for any one observatory site of roughly 1000 hours. Within this time,
the observer must perform two tasks: the observation of the (known or
suspected) gamma-ray source (``on-source'' observation) and an auxiliary 
``off-source'' observation to determine the background caused by hadronic 
cosmic rays contained in the on-source observation.
Different schemes have been developed to perform the ``off-source''
observations. They are described in section \ref{sec-trad}.
All schemes have in common that they either achieve less than optimal
sensitivity  or occupy a large fraction of the
total observation time, roughly 50 \%, which reduces the telescope's
ability to follow the light curves of rapidly variable sources.

In this article, I propose a new method to obtain off-source observations
sacrificing  a smaller fraction of the total observation time 
and avoiding increased systematic uncertainties. Section \ref{sec-trad}
summarizes the presently used background determination methods, section
\ref{sec-new} describes the new method and section \ref{sec-disc} 
discusses advantages and applications.

\section{Traditional background determination}
\label{sec-trad}
In order to better describe the advantages of the proposed new background
measurement method, the methods which have been used so far are briefly
summarized here.

There are essentially two traditional ways of obtaining an 
estimate of the number 
of background events in data from gamma-ray source observations with 
Cherenkov telescopes: true ON/OFF observations and separately taken OFF data. 

\subsection{ON/OFF observations}
This was the first method ever employed in imaging atmospheric Cherenkov
observations \cite{weekes89}.
The background is determined by performing a second observation immediately
before or after the on-source observation (``ON run''). This second 
observation, the ``OFF run'', has the same duration as the ON run
and is made on a celestial position which is the
same as the on-source position except that the Right Ascension is
shifted by the duration of the ON-run increased by the slewing time.

The method achieves a perfect matching of the zenith angle distributions
in ON and OFF run. Also the atmospheric conditions are nearly perfectly
matched since the the runs are taken nearly at the same time.
The only difference between ON and OFF run are (mostly small) variations
in the star field and hence the night sky background noise.
This is to a large extent eliminated by software padding, a method
which uses added noise from a software random generator to equalize
the noise conditions between ON and OFF \cite{cawley}.
Field rotation spreads the star field differences over the
field of view in the course of observations.

Given an observation time (duration of one of the runs) $T$, 
a gamma event rate $R_{\gamma}$ and a background rate $R_b$
(after trigger or arbitrary gamma-hadron separation),
the significance $S$ of the event excess $X = R_{\gamma}T$
caused by the gamma-ray source in the on-source position is
$$
    S = \frac{R_{\gamma} \sqrt{T}}{\sqrt{2 R_b + R_{\gamma}}}
$$

In order to obtain the off-source observations, 50 \%
of the theoretically available on-source observation time has
to be sacrificed. Since some OFF data can be taken while the 
source under investigation is  below the minimum elevation for 
useful observations, the reduction in on-source observation time 
{\it for that particular source} may be somewhat less than 50\%.
But in any case,
determining the hadronic background using ON/OFF observations
decreases the {\it total available} observation time by 50\%.
Furthermore, it introduces  large gaps  in the time coverage
thereby hampering variability studies.

\subsection{Separately taken OFF data} 

Since the hadronic background is known to be isotropic and
time-independent (to a good approximation at energies above
several 10 GeV), it is in principle possible to measure the
background by taking the OFF run long before or long after
the ON run is taken. Also, to decrease the statistical error
of the background measurement, more than one OFF run can
be used. If the analysis is not testing for the presence of a
new (weak) source, the same OFF run can also be used several
times for different ON data. Hence, less than 50\% of the total
observation time has to be sacrificed for OFF data.

If a given observatory were to create a library of OFF source runs
on a grid of all different declinations and zenith angles
of interest, it could
in principle \-- after the completion of the library \-- stop
taking OFF data and re-use the data in the library for all future
analysis (for new source discoveries, new OFF data may have to be
taken for statistical reasons). 

In reality, however, modifications and aging of the
telescope hardware and the ever-changing general atmospheric
conditions cause changes in the characteristics of the data
which make an off-source data library obsolete within a few years \--
roughly the same time than it takes to compile the library.

In case the time coverage for a particular source is to be increased,
exactly matching OFF data can theoretically be taken at a different time
by observing at exactly the same Declination and zenith angle.
This means sacrificing observation time of other sources and
has the additional drawback that,
due to the fact that the atmospheric conditions change on a timescale
of a few hours, it is never possible to reach the near-perfect matching
of the atmospheric conditions  obtained in true ON/OFF
observations. 

To correct for the differences in atmospheric conditions between
ON data and separately taken OFF data, one can use the fraction
of the data at large ALPHA values (which certainly does not come
from the source direction) to normalize the background rate 
$R_b$.\footnote{There are different names and prescriptions for this
method, but all of them are equivalent.}
This can be done in an integral way for the whole gamma signal 
(e.g. \cite{aharonian99b} and references therein, \cite{catanese98}) or
separately for different gamma energy ranges when a spectrum has
to be derived (with or without assuming a correlation between the bins
\cite{petry02}).
In both cases, larger systematic errors on the integral flux
or spectral parameters respectively are the price for not taking
true ON/OFF data. 

\subsection{``Wobble mode''}

If the diameter of the CT's field of view exceeds $4^\circ$, on-source
and off-source observations can be taken at the same time by observing
the known or suspected source position  off-axis by $\approx 0.5^\circ$
and deriving the background
from the analysis of events coming from the mirror position in the other
half of the camera. This so-called
``wobble mode'' (see e.g. \cite{aharonian99a}) gives maximum time coverage
but reduces the effective collection area for gamma-rays by truncating part of the
field of view where the air-shower images from the source direction are expected.
It also shifts the images of the shower maxima of the events of interest into
a region of the field of view which has worse optical properties leading
to a deterioration in the gamma-hadron separation.

The truncation effects become negligible when a camera with a field of
 view larger than
$5^\circ$ becomes available. Equipping telescopes with such large cameras, however,
is often financially impossible. 

In any case, the mirror position in the other half of
the field of view does not provide the exact same zenith angle distribution as
the on-source position leading to additional systematic errors in spectra. 
One way to compensate for this, at least approximately, is to alternate the angle by
which the source position is shifted off-axis between
e.g. $+0.5^\circ$ and $-0.5^\circ$.

\section{The new method}

\label{sec-new}

\subsection{Description}

The new  background determination method proposed here is a 
generalized version of the traditional ON/OFF observations described above.
Instead of having ON and OFF runs of equal duration, the observer
chooses two parameters: the total duration $T_t$ of the two runs together and 
the fraction $f$ of $T_t$ which is used for the OFF observation.
The ON observation time is then
\begin{equation}
   T_{\mathrm{on}} = (1 - f) T_t
\end{equation}
while the OFF observation time is
\begin{equation}
   T_{\mathrm{off}} = f T_t
\end{equation}
The ON observation is then performed with duration $T_{\mathrm{on}}$ as
usual. The OFF observation, however, since it has in the general case a 
duration different from the ON observation,  has to be performed
at a {\it different tracking speed} in order to cover the same
zenith angle range. The tracking speed is scaled by the
ratio of the run durations $c$ where
\begin{equation}
    c = \frac{1-f}{f}
\end{equation}
The correction of the tracking speed can simply be achieved by
{\it substituting the absolute time $t$ in the tracking calculations
by a modified absolute time $t'$} given by
\begin{equation}
   t' = t_{\mathrm{start}} + c \cdot (t - t_{\mathrm{start}})
\end{equation}
where $t_{\mathrm{start}}$ is the (unmodified) absolute
time at the beginning of the OFF run.

\begin{figure}
\leavevmode
\centering
\epsfxsize=15cm
\epsffile{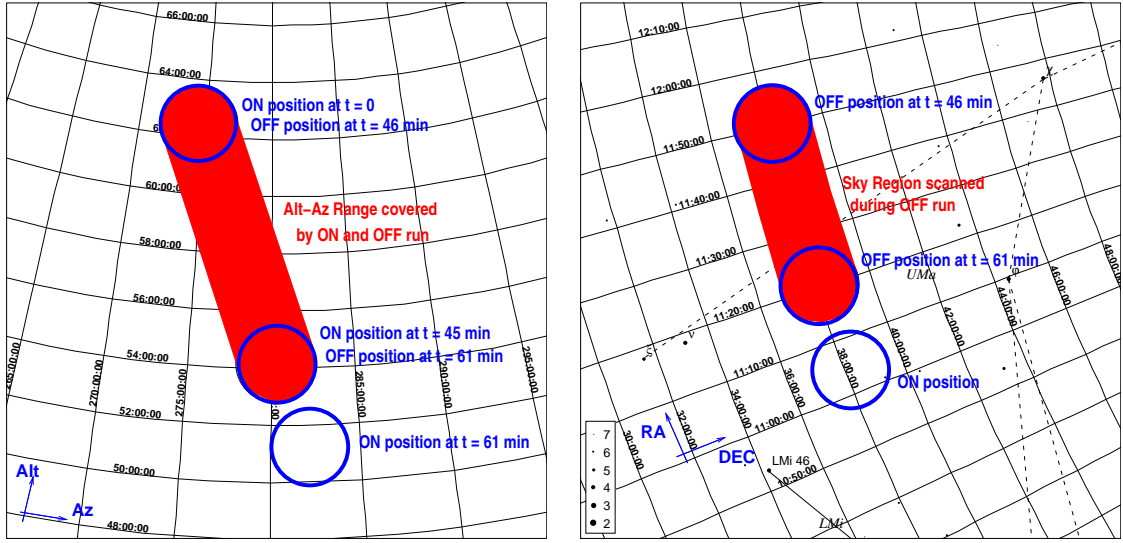}
\caption{\label{fig-example} Example of the application of FBD to an
observation of the Blazar Mkn 421 with $T_t = 60$~min, $f = 0.25$
($\Rightarrow T_{\mathrm{on}} = 45$~min, $T_{\mathrm{off}} = 15$~min),
slewing time between runs = 1~min. {\bf Left:} The ON and OFF run seen in the 
local Alt-Az coordinate system. {\bf Right:} The same seen on the RA-DEC
coordinate system. The circles indicate the typical telescope acceptance 
for gamma-like airshowers and have 2.4$^\circ$ diameter.
}
\end{figure}

This substitution has the effect that if $f \neq 0.5$, the tracking speed
is faster ($f < 0.5$) or slower ($f > 0.5$) than normal.
For example, if $f$ is chosen to be $0.25$ and $T_t = 60$~minutes,
then $T_{\mathrm{on}} = 45$~minutes, $T_{\mathrm{off}} = 15$~minutes
and $c = 45/15 = 3$. In this case, the telescope would be tracking
three times faster during the OFF run.
Figure \ref{fig-example} illustrates this example.

{\it The RA/DEC coordinates used for the OFF run are the same
as for the traditional ON/OFF case.}
If $f$ is chosen to be $0.5$, the observation is a traditional
ON/OFF observation.

Given the number of events after arbitrary analysis stages for
the ON and OFF run, $N_\mathrm{on}$ and $N_\mathrm{off}$,
the number of excess events (``gammas'') $X$ is calculated
as
\begin{equation}
    X = N_\mathrm{on} - c  N_\mathrm{off}
\end{equation}
and the significance $S$ of this signal is
\begin{equation}
\label{equ-sig1}
    S = \frac{X}{\Delta X} = 
           \frac{ N_\mathrm{on} - c  N_\mathrm{off}}{\sqrt{N_\mathrm{on} + c^2 N_\mathrm{off}}}
\end{equation}
The error of $c$ is negligible since it is implemented
by a comparatively very accurate time measurement.

As will be shown further below, $f = 1/(c+1)$ should always be chosen
to be $\geq 0.5$. The modified time $t'$ is therefore always
faster than normal time. Hence the new method is named
{\it Faster Background Determination} (FBD).

{\it Note that the fact that the telescope is not tracking a fixed point
in the sky during the OFF run (because it is moving faster than
the Earth's rotation) does not compromise the data quality.
Field rotation leads to a changing starfield configuration anyway,
also for the traditional ON/OFF case. A superimposed drift of the
starfield (a few degrees within 15 minutes in the typical case) will
not change this situation\footnote{Modern approaches to dealing with bright
stars in the field of view either take the affected photomultiplier tubes out 
of the trigger logic or lower the high voltage on them but do not switch them
off. This happens in a computer-controlled, reproducible fashion. Therefore
there is no additional new precaution necessary to deal with a 
drifting starfield as opposed to a purely rotating one. The drift of 
the OFF starfield leads to an increase of the probability of having a
bright star in the field of view by about a factor up to 3. See the
discussion in section \protect\ref{sec-disc}.}.}

Depending on the angular diameter $d_{\gamma}$ of the part of the
telescope's field of view from which gamma-like shower images are
accepted in the data analysis, 
a {\it minimum
OFF run duration has to be required} to avoid overlap of the ON
and OFF regions. The minimum duration $T_{\mathrm{off,min}}$ is given by
\begin{equation}
\label{equ-fov}
   T_{\mathrm{off,min}} = \frac{d_{\gamma}}{\cos(\mathrm{DEC})} \cdot 4\ \mathrm{minutes/degree}
\end{equation}
where $\mathrm{DEC}$ is the Declination of the ON source position.
For a typical CT with $d_{\gamma} = 2.4^\circ$, $T_{\mathrm{off,min}}$
would be 9.6~min$/\cos(\mathrm{DEC})$, i.e. between 9.6~min and 15~ min for
 $|\mathrm{DEC}| < 50^\circ$.

\subsection{Optimization}

The new method (FBD) has two parameters which have to be chosen
by the observer: total observation time $T_t$ for one ON/OFF run pair
and the time fraction $f$ used for the OFF run. 

The choice of $T_t$ is dictated by practical considerations,
the timescale of changes in the atmospheric conditions and
the field of view of the camera (Equation \ref{equ-fov}).

\begin{figure}[t]
\leavevmode
\centering
\epsfxsize=11.5cm
\epsffile{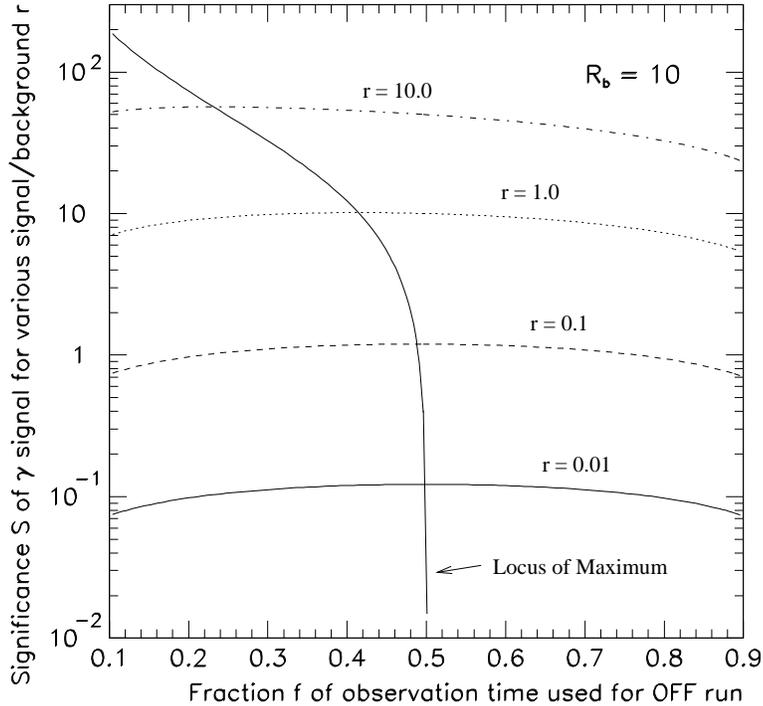}
\caption{\label{fig-sigvsf} \small Assuming a background rate $R_b = 10$
 and a run pair duration $T_t = 60$ (both in arbitrary units), the
significance $S$ of a gamma signal with gamma rate to background ratio 
$r = R_{\gamma}/R_b = 0.01, 0.1, 1.0, 10.0$ is plotted versus
the OFF-run time fraction $f$. The curve labeled ``Locus of Maximum''
connects the maxima of all significance curves.}
\end{figure}

The optimal choice of $f$ is obviously independent of $T_t$,
but it depends \-- as it turns out \-- on the intensity of the
gamma-ray source. In order to see this, Equation \ref{equ-sig1}
for the significance $S$ of the gamma signal is rewritten  
substituting 
$$
  c = \frac{1-f}{f},\  N_\mathrm{on} =  (R_\gamma + R_b)(1-f)T_t,\ 
  N_\mathrm{off} =   R_b f T_t
$$
where $R_\gamma$ is the gamma event rate and $R_b$ is the
background event rate.
This gives
\begin{equation}
S = \frac{(R_\gamma + R_b)(1-f)T_t - R_b f T_t}
          {\sqrt{(R_\gamma + R_b)(1-f)T_t + (\frac{1-f}{f})^2 R_b f T_t}}\\
  = \frac{R_{\gamma}\sqrt{(1-f)T_t}}{\sqrt{R_\gamma + R_b + \frac{1-f}{f} R_b}}
\end{equation}
Choosing example values for $R_\gamma$, $R_b$ and $T_t$ , the 
significance $S$ can be plotted versus $f$ in order to investigate the
dependence. This was done in Figure \ref{fig-sigvsf} for $T_t = 60$ with
$R_b = 10$ and $r = R_{\gamma}/R_b = 0.01, 0.1, 1.0, 10.0$.

The remarkable result which becomes visible in Figure \ref{fig-sigvsf}
is that {\it as the signal to background ratio increases, the position $f_0$ 
of the maximum in the significance curve decreases}.

\begin{figure}[b]
\leavevmode
\centering
\epsfxsize=11.5cm
\epsffile{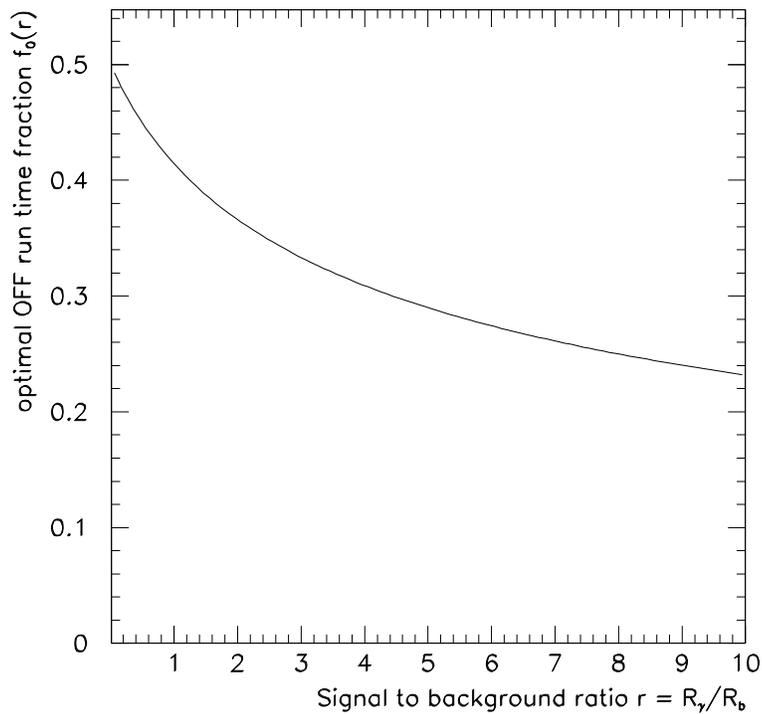}
\caption{\label{fig-f0vsr} \small The optimal value of $f$ where maximum
gamma signal significance is obtained as a function of the ratio $r$
of gamma event rate and background event rate (see Equation \protect\ref{equ-f0}).
}
\end{figure}

The value $f_0$ for which the maximum significance is obtained, can be
analytically calculated by determining the zeros of the derivative
d$S$/d$f$:
\begin{equation}
\frac{\mathrm{d}S}{\mathrm{d}f} = 
 \frac{R_\gamma ( R_b + 2 R_b f + f^2 R_\gamma ) T_t}
     {2 f \sqrt{R_b/f + R_\gamma} ( R_b + f R_\gamma ) \sqrt{(1-f)T_t}} = 0
\end{equation}
This is essentially a quadratic equation in f. One finds that one of the zeros
is always negative and therefore not physical in this context. The
remaining zero is
\begin{eqnarray}
f_0 &=& \sqrt{(R_b/R_\gamma)^2 + R_b/R_\gamma} - R_b/R_\gamma\\
\label{equ-f0}
    &=& \sqrt{\frac{1}{r^2} + \frac{1}{r}} - \frac{1}{r}
\end{eqnarray}
where $r = R_{\gamma}/R_b$ is the signal/background ratio
(not signal/noise!) as above. Figure \ref{fig-f0vsr} shows $f_0$ as a function of $r$.
The figure and Equation \ref{equ-f0} show an interesting property of the
FBD scheme:

{\it For each observation, there is a single optimal value of $f$ which 
  depends only on the ratio $r$ of the gamma event rate
  and the background event rate. This value is always less than 0.5
  but approaches 0.5 asymptotically with decreasing $r$.}
     
In other words, the traditional value $f = 0.5$ is only optimal for
   weak gamma sources. For stronger sources like the Blazar Mkn 421 during
   a flare, one obtains a more significant signal if one
   devotes more time to the ON than to the OFF observation,
   i.e. chooses $f<0.5$.

\section{Discussion}

\label{sec-disc}

The Faster Background Determination method will bring three major advantages:

\begin{enumerate}

\item Increased time coverage without increased systematic errors
  in the determination of flux and spectrum compared to normal ON/OFF 
  observations.

\item Reduction of systematic errors and simplification of data analysis
    compared to methods using separately taken OFF data.

\item Moderate improvement of the statistical accuracy of flux and spectral 
   measurements since the significance of the gamma signals is maximized for 
   a given total observation time.

\end{enumerate}
The most important are points 1 and 2.

Concerning point 3, one can show that the maximum possible increase in significance 
when using FBD instead
of normal ON/OFF observations is only dependent on the signal to
background ratio $r = R_{\gamma}/R_b$ and is described by
the following formula:
\begin{equation}
\label{equ-sincvsr}
\frac{S_{\mathrm{FBD}}}{S_{\mathrm{norm}}}
  = \sqrt{2 (1 + \frac{2}{r}) \frac{1 + r - \sqrt{r + 1}}{1 + r + \sqrt{r + 1}}}
\end{equation}
where $S_{\mathrm{FBD}}$ is the significance obtained by making an
ON/OFF observation in FBD mode using the optimal OFF time fraction $f_0$
(Equation \ref{equ-f0}) and $S_{\mathrm{norm}}$ is the significance
when the OFF time fraction $f = 0.5$ is used instead (normal ON/OFF run). 

\begin{figure}[htb]
\leavevmode
\centering
\epsfxsize=10cm
\epsffile{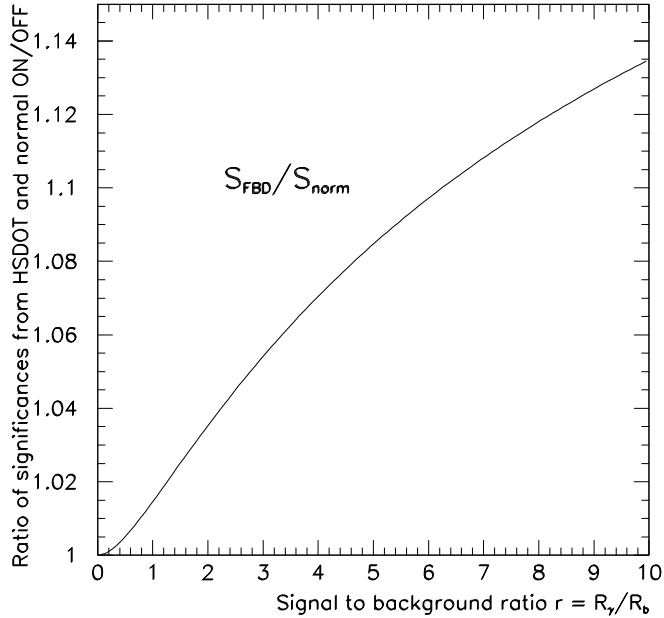}
\caption{\label{fig-sincvsr} \small The improvement of the gamma signal significance
when using FBD instead of normal ON/OFF runs. (See Equation \protect\ref{equ-sincvsr}.)
}
\end{figure}

Figure \ref{fig-sincvsr} shows the ratio described by Equation \ref{equ-sincvsr}.
It is always larger than unity, i.e. FBD is always better than normal ON/OFF
if the optimal OFF time fraction $f_0$ is used. 

\begin{table}[hb]
\caption{\label{tab-comp} \small Properties of the background determination
methods under discussion.}
{\tiny
\begin{tabular}{l|c|c|c|c}
          & Traditional ON/OFF & Separate OFF & Wobble$^*$ & FBD \\
\hline
Max. possible & $\approx$ 60\% & 100\% & 100\% & $\approx$ 80\% \\ 
\ one-source time & & & \\
\ coverage & & & \\
\hline
Max. possible &   50\% & $\approx$ 66\% & 100\% & $\approx$ 80\% \\ 
\ all-source time & & & \\
\ coverage & & & \\
\hline
Statistical Errors & standard & smaller & smaller & smaller \\
\ for same overall &          & by $\leq$ 30 \%  & by $\leq$ 30 \% & by $\leq$ 15 \%\\
\ obs. time  & & & \\
\hline
Systematic Errors & minimal &  larger  & larger &  minimal \\
\hline
Application      &  precision  & new source & new source  & precision \\
                  &  measurements & search   & search,    & measurements \\
                  & of weak sources &        & multi-$\lambda$ & any source strength, \\
                  &                 &          &  campaigns   & multi-$\lambda$ campaigns \\
\hline
\end{tabular}

$^*$only possible if diameter of camera field of view included in trigger $> 4^\circ$
}
\end{table}

Of course, for unknown sources, the signal to background ratio is unknown and
hence $f_0$ cannot be determined. However, as one can see from Figure \ref{fig-sigvsf},
the maxima of the significance curves are broad and an approximate value
for $f$ already gives good results. 
This also means that using FBD, the time coverage can be increased significantly
in exchange for only a small decrease in sensitivity.

Generally, when in discovery mode where accuracy of flux measurements is not
the primary concern,  observations with separately taken OFF data or
wobble mode observations may be an 
equally good way to find a new source. But as soon as the presence of the source is
established and accurate flux and spectral measurements are of interest,
FBD is the method of choice. 

Due to the increase in the sky area covered by the OFF observation using FBD,
the likelihood that bright stars occur in the OFF region increases
by a factor up to $\approx$ 3 depending on the choice of the OFF 
run time fraction $f$. Figure \ref{fig-example} already 
shows the mildly extreme case with $f=0.25$. Near the 
galactic plane, CTs have traditionally
had problems with their background determination due to the 
presence of many bright stars. The FBD
method will only slightly worsen an already difficult problem.
CTs have to work with low photomultiplier gain and the above mentioned
dynamical lowering of high voltage values and modification of the 
trigger map to make progress here. 

The value of the signal to background ratio $r$ depends on the gamma-hadron
separation capability of the telescope and the state of the source.
FBD seems to be particularly helpful for the observation of Blazars
because (a) they  reach the highest values of $r$, (b) they have
unproblematic starfields surrounding them as most of them are sufficiently far
away from the galactic plane, and (c) due to their variability, time-coverage
is of interest.
The observation schedules of all CT observatories have always made 
a special effort to dedicate large fractions of the observation time to
flaring Blazars. For example, in 1997, when Mkn 501 showed an unprecedented
flaring state of several months duration, observatories were dedicating
more than 50 \% of the available time to this source.

Finally, for very much the same reasons that FBD is beneficial for
Blazar observations, it will also be beneficial for gamma-ray burst 
follow-up. Choosing $f$ very low (possibly as low as $0.1$), it will be 
possible to maximize the on-source time without compromising the quality 
of the background determination.
 
Table \ref{tab-comp} summarizes the properties of FBD and the traditional 
background determination methods.

\end{document}